\documentclass[english,aps,manuscript,aps,manuscript,showpacs]{revtex4}
\usepackage[T1]{fontenc}
\usepackage[latin9]{inputenc}
\usepackage{float}
\usepackage{amstext}
\usepackage{amssymb}
\usepackage{graphicx}

\makeatletter
\@ifundefined{textcolor}{}
{%
 \definecolor{BLACK}{gray}{0}
 \definecolor{WHITE}{gray}{1}
 \definecolor{RED}{rgb}{1,0,0}
 \definecolor{GREEN}{rgb}{0,1,0}
 \definecolor{BLUE}{rgb}{0,0,1}
 \definecolor{CYAN}{cmyk}{1,0,0,0}
 \definecolor{MAGENTA}{cmyk}{0,1,0,0}
 \definecolor{YELLOW}{cmyk}{0,0,1,0}
}

\makeatother

\usepackage{babel}
\begin{document}

\title{Controlled interference of association paths in the conversion of
ultracold atoms into molecules }

\author{J. Plata}

\affiliation{Departamento de F\'{\i}sica, Universidad de La Laguna,\\
 La Laguna E38204, Tenerife, Spain.}
\begin{abstract}
We present a proposal for controlling the conversion of ultracold
atoms into molecules by fixing the phase difference between two oscillating
magnetic fields. The scheme is based on the use of a magnetic Feshbach
resonance with a field modulation that incorporates terms oscillating
with frequencies corresponding to the main resonance and one of the
subharmonics. The interference between the two association processes
activated by the oscillating terms is controlled via the phase difference.
As a result, significant increase or decrease of the effective interaction
strength can be achieved. The realization of the proposal is feasible
under standard technical conditions. In particular, the method is
found to be robust against the effect of the sources of decoherence
present in the practical setup. The applicability of the approach
to deal with quadratic terms in the field modulation is discussed.
\end{abstract}


\maketitle

\section{Introduction}

The use of Feshbach resonances (FRs) has been crucial for the advances
made in the last years in the study of ultracold atoms and molecules
\cite{key-ReviewKohler,key-Timmermans,key-Donley,key-Jin1,key-Hulet1,key-Greiner1,key-Mukaiyama1,key-Cubizolles,key-Grimm1}.
The control of the position and width of a FR, and, consequently,
the manipulation of the dynamics of the atom-molecule system, have
been the objectives of intense work. In this line, interesting proposals,
which incorporate the use of different fields and transitions, have
been made \cite{key-DalibardControl,key-KokkelControl,key-Rempe-Control,key-KaufmanControl}.
These control issues are particularly relevant to the optimization
of the methods employed in the production of ultracold molecules.
(The progress in this area is continuous; in this sense, it is worth
pointing out interesting experimental work reported in \cite{key-klempt,key-lange,key-beaufils,key-takekoshi}.)
One of the techniques frequently applied to generate ultracold molecules
from both Fermi and Bose atomic gases is based on a sinusoidal modulation
of the field in a magnetic FR \cite{key-ThomsonAssoc,key-Hanna Assoc,key-Inguscio,key-HuletSub,key-BrouardAssoc,key-Braaten}.
With this arrangement, molecule formation has been observed not only
at the main resonance, i.e., for a field frequency matching the molecular
binding energy, but also at subharmonic resonances. Here, we propose
a variation of this scheme that introduces additional elements of
control in the conversion process. Specifically, we consider a magnetic
modulation that incorporates two terms oscillating with different
frequencies. In the particular case proposed to illustrate the method,
one of the frequencies corresponds to the main resonance and the other
to one of the subharmonics. The phase difference between the driving
terms can be used to steer the interference between the processes
generated via the two considered resonances, and, in turn, to optimize
(or inhibit) the production of molecules. Our approach to the system,
based on the use of appropriate unitary transformations to simplify
the description, allows an analytical characterization of the transition
process, and, consequently, a direct application of the results to
the implementation of strategies of control. The robustness of the
scheme against the dephasing mechanisms potentially present in the
considered scenario is evaluated. The practical realization of the
interference paths, viable under usual experimental conditions, can
serve to assess our understanding   of the subharmonics resonances.
It can also provide a valuable test on the assumed coherent response
of the atom-molecule system. The general components of the developed
method, combined in different forms, can found applicability in various
contexts. In this sense, differential implications of variations of
the basic scheme are discussed. 

The outline of the paper is as follows. In Sec. II, a first general
description of basic two-body dynamical characteristics of the system
is presented. The specific proposal for the interference scheme is
introduced in Sec. III. Additionally, the applicability of the theoretical
framework to account for second-order effects in the modulation field
is assessed. In Sec. IV, we analyze many-body aspects of the dynamics
 and discuss the relevance of the different sources of decoherence
present in the system realization to the validity of the method. Finally,
some general conclusions are summarized in Sec. V.

\section{Two-body dynamics}

We consider a Bose gas of ultracold atoms in a weakly confining harmonic
trap. In order to induce molecule association, a magnetic FR is applied:
the variation of the magnetic field alters the detuning between the
(closed-channel) FR state $\left|R\right\rangle $ and the (entrance-channel)
atom state $\left|S\right\rangle $, and, consequently, modifies the
effectiveness of the inter-channel coupling. This opens the possibility
of manipulating the scattering length. Moreover, depending on its
functional form, the time-dependent field can induce transitions between
the entrance states and the bound state. This second implication will
be crucial for the scheme proposed in our study. (As in the original
FR scenario, we consider that the interchannel coupling is due to
hyperfine interactions; the generalization of our approach to deal
with  modifications of the coupling is direct.) Hence, in the considered
arrangement, the magnetic field is modulated as 

\begin{equation}
B(t)=B_{0}+B_{m}(t),
\end{equation}
where the average value $B_{0}$ is assumed to be close to a FR position
$B_{r}$. Additionally, in order to work in a perturbative regime,
we consider that the modulation field $B_{m}$ is sufficiently small,
$\left|B_{m}\right|\ll\left|B_{0}-B_{r}\right|$.  {[}The proposed
functional form of $B_{m}(t)$, appropriate for our strategies of
control, will be specified further on.{]} The weak-confinement conditions
allow us to apply a local-density approximation. In our procedure,
we focus on the simplest case of a uniform system. Consistently, the
atomic energy levels, which are closely spaced because of the weak
trapping, will be approximated as a quasi-continuum. The robustness
of the approach against the effect of nonuniformities in the field
will be discussed.

\subsection{The undriven system: the bare and the dressed-state representations}

In the absence of magnetic modulation, (i.e., for $B_{m}=\textrm{0}$),
the Hamiltonian reads $H_{1}=H_{0}+V$, where $H_{0}$ stands for
the Hamiltonian of the interaction-less system at the mean magnetic
field $B_{0}$, and $V$ represents the coupling term between the
eigenstates of $H_{0}$, i.e., the \textit{bare states}, $\left|R\right\rangle $
and $\left|S\right\rangle $. Specifically, we have \cite{key-BrouardAssoc}

\begin{eqnarray}
H_{0}\left|S\right\rangle  & = & E_{S}\left|S\right\rangle ,\;\;\;\;\;0\leq E_{S}<\infty,\nonumber \\
H_{0}\left|R\right\rangle  & = & -\epsilon_{B_{0}}\left|R\right\rangle ,\;\;\;\;\;\epsilon_{B_{0}}>0,\nonumber \\
\left\langle R\right|V\left|S\right\rangle  & = & v(E_{S}),
\end{eqnarray}
where $\epsilon_{B_{0}}$ denotes the binding energy of the state
$\left|R\right\rangle $ at the field $B_{0}$, and $E_{S}=p^{2}/m$
represents the energy along the atomic set, ($p$ is the relative
linear momentum and $m$ stands for the atom mass.) The operator $V$,
which accounts for hyperfine interaction between the atom pairs in
the open and in the closed channels, affects only nuclear and electron
spin variables, which are fixed in each channel. Therefore, the dependence
of the interaction strength $v(E_{S})$ on the atom-state characteristics
is determined by the overlap between the wave functions $\psi_{R}(\vec{r})=\left\langle \vec{r}\right|\left.R\right\rangle $
and $\psi_{S}^{(p)}(\vec{r})=\left\langle \vec{r}\right|\left.S\right\rangle $,
where $\vec{r}$ denotes the relative coordinate. Then, the functional
form of $v(E_{S})$ can be determined through the evaluation of the
Franck-Condon factor

\begin{equation}
F(p)\equiv\left|\int d\vec{r}\psi_{R}^{\star}(\vec{r})\psi_{S}^{(p)}(\vec{r})\right|^{2}.
\end{equation}
In the considered range of energies, $v(E_{S})$ can be assumed to
hardly vary with $E_{S}$. Moreover, we assume that the set of \textit{dressed
states} of the system, (i.e., of eigenstates of the complete Hamiltonian
$H_{1}$), is formed by a bound eigenstate $\left|M\right\rangle $,
(the molecular state), with energy $-\epsilon_{M}$ , and by a continuum
of states $\left|A\right\rangle $ with energy $E_{A}$. Namely, 

\begin{eqnarray}
H_{1}\left|A\right\rangle  & = & E_{A}\left|A\right\rangle ,\;\;\;\;\;0\leq E_{A}<\infty,\nonumber \\
H_{1}\left|M\right\rangle  & = & -\epsilon_{M}\left|M\right\rangle ,\;\;\;\;\;\epsilon_{M}>0.
\end{eqnarray}
As the field value $B_{0}$ approaches the FR position $B_{r}$, the
dressed and the bare-state representations increasingly differ. In
particular, $\left|M\right\rangle $ departs from $\left|R\right\rangle $.
Hence, in the considered situations, which require the description
of atom-molecule transfer processes near $B_{r}$, the use of the
dressed-state basis is necessary.

\subsection{The driven dynamics in the dressed-state representation}

We assume that, as in standard practical setups \cite{key-ThomsonAssoc},
the time-dependent magnetic field is applied through a \emph{trapezoidal}
ramp similar to that outlined in Fig. 1. In the first linear segment,
the field is slowly varied from $B(t=t_{i})$, (far from the FR),
to $B_{0}$. In the central \emph{plateau}, the modulation $B_{m}(t)$
introduced in Eq. (1) is connected. Subsequently, a slow linear ramp
drives $B(t)$ back to the initial value. A detailed study of the
incorporation of this type of arrangements in the theoretical description
was presented in \cite{key-BrouardAssoc}. The methodology applied
there combines the following elements. First, the effects of the (slow)
linear segments of the ramp are simulated through an adiabatic approximation
that connects the bare and dressed representations. Second, to deal
with the dynamics in the central region, a dressed-state approach
is implemented. The adiabatic segments can be regarded as part of
the system preparation. (Their role in different practical setups,
, i.e., in arrangements corresponding to a condensate preparation
and to an initial thermal mixture, was analyzed in \cite{key-BrouardAssoc}.)
Since, that analysis is applicable to the present context, we will
concentrate here on the differential aspects introduced by the considered
modulation in the dynamics in the central \emph{plateau}.

\bigskip{}

\includegraphics[scale=0.4]{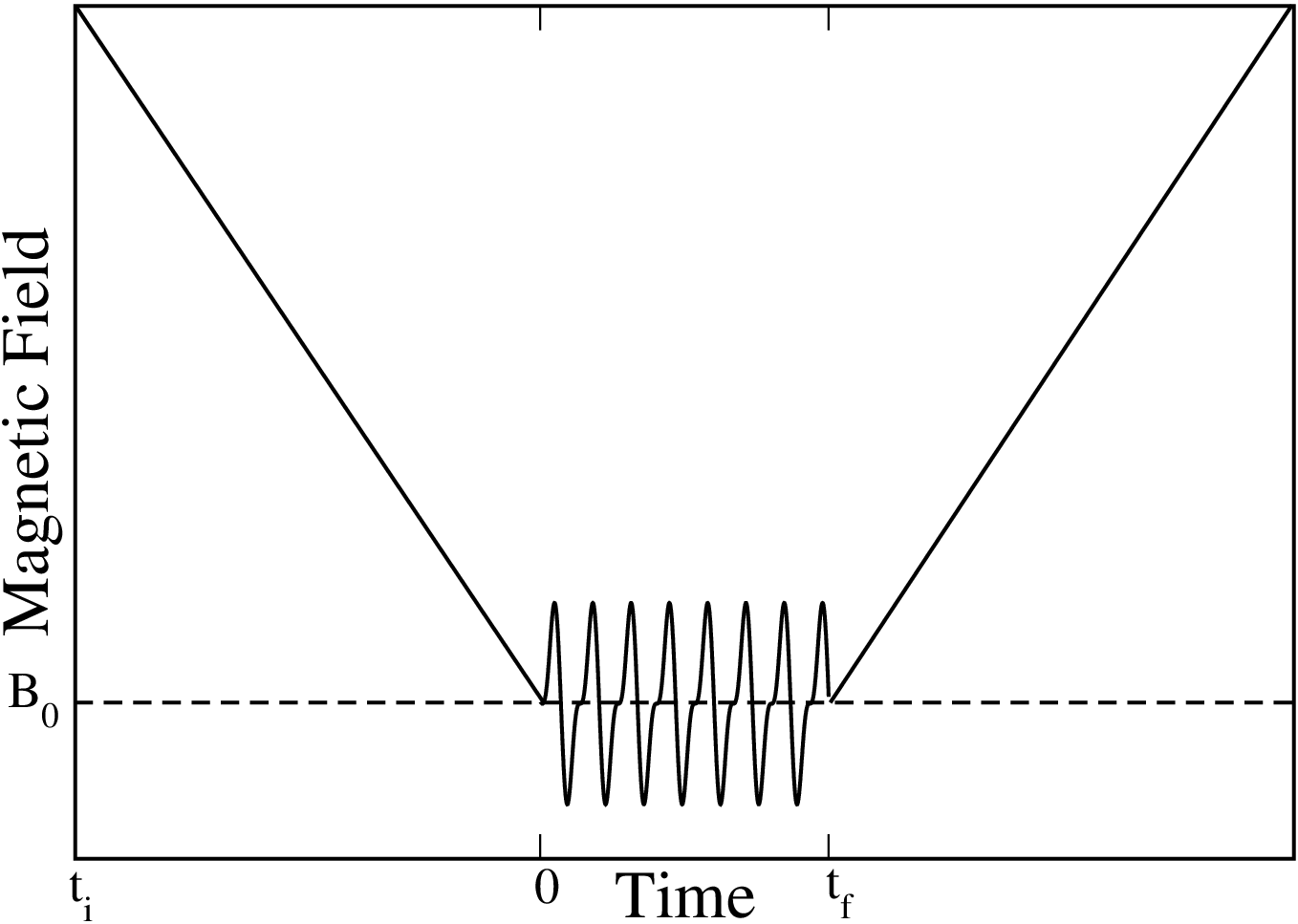}

\begin{figure}[H]
\caption{A diagram (with arbitrary scales) of the magnetic-field ramp.}
\end{figure}

The incorporation of the magnetic modulation into our theoretical
framework is straightforward in the bare-state picture: the field
variation alters the binding energy of the FR state according to $\epsilon_{B_{0}}\rightarrow\epsilon_{B_{0}}+\hbar C_{B_{0}}B_{m}(t)$,
where $\hbar C_{B_{0}}=\left[\frac{\partial\epsilon}{\partial B}\right]_{B_{0}}$
corresponds to the difference between the magnetic moments of the
involved states. (In the following, we will take $\hbar=\textrm{1}$.)
(The effect of field-quadratic corrections to the binding energy will
be considered later on.) As a consequence, the complete system is
described by the  Hamiltonian $H_{2}=H_{1}+C_{B_{0}}B_{m}(t)\left|R\right\rangle \left\langle R\right|$,
which, in the dressed-state basis has the form

\begin{eqnarray}
H_{2} & = & \left[-\epsilon_{M}+\eta_{M}B_{m}(t)\right]\left|M\right\rangle \left\langle M\right|+\left[E_{A}+\eta_{A}(E_{A})B_{m}(t)\right]\left|A\right\rangle \left\langle A\right|+\nonumber \\
 &  & \left[v{}_{eff}(E_{A})B_{m}(t)\left|M\right\rangle \left\langle A\right|+\textrm{h.c.}\right],
\end{eqnarray}
where

\begin{eqnarray}
\eta{}_{M} & = & C_{B_{0}}\left|\left\langle M\right|\left.R\right\rangle \right|^{2}\\
\eta_{A}(E_{A}) & = & C_{B_{0}}\left|\left\langle A\right|\left.R\right\rangle \right|^{2}\\
v_{eff}(E_{A}) & = & C_{B_{0}}\left\langle M\right|\left.R\right\rangle \left\langle R\right|\left.A\right\rangle .
\end{eqnarray}
Notice that the field modulation has a twofold effect: it introduces
a time-dependent coupling between the states $\left|M\right\rangle $
and $\left|A\right\rangle $, and induces a time variation in their
\emph{eigenenergies}. Although the modulation can also couple different
dressed atomic states, that interaction has been neglected due to
the small magnitude of the involved state-projections, and, given
that, in the considered situations, the atomic energy spacing is far
from being in resonance with the driving field.

\section{The interference scheme}

\subsection{General approach}

Our proposal for manipulating the dynamics is based on the application
of a magnetic-field modulation $B_{m}(t)$ composed of two oscillating
terms with different frequencies and amplitudes and a constant phase
difference. (The generalization to more than two terms is direct.)
Namely, in the central region of the previously indicated trapezoidal
ramp, we connect a modulation field with the form

\begin{equation}
B_{m}(t)=B_{1}\sin(\omega_{1}t)+B_{2}\sin(\omega_{2}t+\chi).
\end{equation}
To tackle the resulting dynamics, we first apply the unitary transformation
\begin{equation}
U(t)=e^{i\left[\left(\omega t+\eta_{M}F(t)\right)\left|M\right\rangle \left\langle M\right|+\eta_{A}F(t)\left|A\right\rangle \left\langle A\right|\right]}
\end{equation}
where we have introduced the function

\begin{equation}
F(t)=\frac{B_{1}}{\omega_{1}}\cos(\omega_{1}t)+\frac{B_{2}}{\omega_{2}}\cos(\omega_{2}t+\chi).
\end{equation}
{[}The value of $\omega$ will be chosen, depending on the particular
(frequency) characteristics of the scheme, to simplify the picture
brought about by the application of $U(t)$.{]} As it corresponds
to a time-dependent unitary transformation \cite{key-castin}, the
transformed Hamiltonian is given by $H_{2}^{\prime}=U^{\dagger}H_{2}U-iU^{\dagger}\dot{U}$.
It reads 

\begin{eqnarray}
H_{2}^{\prime} & = & (-\epsilon_{M}+\omega)\left|M\right\rangle \left\langle M\right|+E_{A}\left|A\right\rangle \left\langle A\right|+\nonumber \\
 &  & B_{m}(t)\left[v_{eff}e^{-i\left[\omega t+(\eta_{M}-\eta_{A})F(t)\right]}\left|M\right\rangle \left\langle A\right|+\textrm{h.c.}\right].
\end{eqnarray}
Now, for the terms that contain $F(t)$, we take into account the
expansion of the exponential of a sine or cosine function in terms
of the Bessel functions $J_{k}(x)$ \cite{key-Grad}, and rewrite
the Hamiltonian as 

\begin{eqnarray}
H_{2}^{\prime} & = & (-\epsilon_{M}+\omega)\left|M\right\rangle \left\langle M\right|+E_{A}\left|A\right\rangle \left\langle A\right|+\nonumber \\
 &  & \left[v_{eff}\left(B_{1}G_{1}(t)+B_{2}G_{2}(t)\right)\left|M\right\rangle \left\langle A\right|+\textrm{h.c.}\right],
\end{eqnarray}
where we have incorporated the functions

\begin{eqnarray}
G_{j}(t) & = & \frac{1}{2}\sum_{m=-\infty}^{\infty}\sum_{q=-\infty}^{\infty}(-i)^{m+q+1}J_{m}(\zeta_{1}B_{1})J_{q}(\zeta_{2}B_{2})\times\nonumber \\
 &  & e^{-i\left[(\omega+m\omega_{1}+q\omega_{2})t+q\chi\right]}\left(e^{i\omega_{j}t}-e^{-i\omega_{j}t}\right),\;(j=1,2),
\end{eqnarray}
and the coefficients

\begin{equation}
\zeta_{j}=\frac{\eta_{M}-\eta_{A}}{\omega_{j}},\;(j=1,2).
\end{equation}
The form of $H_{2}^{\prime}$ given by Eq. (13) {[}along with Eqs.
(14) and (15){]} provides a framework for simplifying the description.
Specifically, an analytical coarse-grained picture of the dynamics
can be derived through the averaging of $H_{2}^{\prime}$. Apart from
the diagonal term, (where the effective detuning $-\epsilon_{M}+\omega$
will be minimized by the appropriate choice of $\omega$), $H_{2}^{\prime}$
has a series of interaction terms oscillating with frequencies determined
by $\omega$, $\omega_{1}$, $\omega_{2}$, and by the integer numbers
$m$ and $q$. The secular dynamics is then governed by the components
with zero frequencies. The rest of terms  barely affect the system
evolution: they present fast oscillations, which, because of the small
magnitude of their amplitudes compared with the involved frequencies,
can be averaged out to zero. (We recall that a perturbative regime
is considered for the modulation amplitudes.) The secular terms can
be easily identified if $\omega$ is appropriately chosen as a function
of the characteristic frequencies of the driving. Additionally, for
the previous arguments on the negligible role of the highly oscillating
terms to be applicable, the choice of $\omega$ must lead to a significant
reduction of the effective detuning. This general procedure, which
is applicable to different combinations of frequencies and amplitudes
of the driving field, will be particularized, in the following, to
a specific proposal.

\subsection{Interference of the association paths corresponding to the main resonance
and the first subharmonic}

Let us consider a particular arrangement where the frequencies of
the driving field are $\omega_{1}=2\omega_{2}\simeq\epsilon_{M}$.
Then, the appropriate value of the characteristic frequency $\omega$,
to be introduced in Eq. (11), is $\omega=\omega_{1}$: this choice
minimizes the effective detuning and allows us to straightforwardly
single out the secular terms in Eq. (14). We can go further by identifying
the dominant contribution among those terms. Indeed, taking into account
the perturbative character of the modulation field, the arguments
of the Bessel functions in Eq. (14) can be assumed to be small enough
for the magnitude of those functions to significantly decrease as
the absolute value of the integer index grows. Hence, the dominant
stationary term in $G_{1}(t)$ corresponds to the indexes $m=0,\: q=0$.
In parallel, in $G_{2}(t)$, the most important secular contribution
comes from the term with $m=0,\: q=-1$. (In order to simplify the
presentation, we focus on the effect of the first-order contributions.
The inclusion of the higher-order corrections, albeit lengthy, is
direct.) Therefore, in the optimum frequency range, namely, for $\omega=\omega_{1}\sim\epsilon_{M}$,
the dynamics is approximately described by the reduced Hamiltonian 

\begin{eqnarray}
H_{eff} & = & E_{M}\left|M\right\rangle \left\langle M\right|+E_{A}\left|A\right\rangle \left\langle A\right|+\nonumber \\
 &  & \left[\tilde{v}_{eff}(E_{A})\left|M\right\rangle \left\langle A\right|+\textrm{h.c.}\right],
\end{eqnarray}
where the effective energy of the molecular state is given by $E_{M}\equiv\omega_{1}-\epsilon_{M}$,
and

\begin{equation}
\tilde{v}_{eff}\equiv-J_{0}(\zeta_{1}B_{1})\left[iJ_{0}(\zeta_{2}B_{2})B_{1}+J_{1}(\zeta_{2}B_{2})e^{i\chi}B_{2}\right]\frac{v_{eff}}{2}
\end{equation}
is a \textit{renormalized} coupling constant which incorporates the
magnetic modulation. As the splitting between the effective atomic
and molecular levels reads $\Delta\equiv E_{M}-E_{A}=\omega_{1}-(\epsilon_{M}+E_{A})$,
the resonance frequency is given by $\omega_{1}^{(R)}=\epsilon_{M}+E_{A}$.
From these results, some preliminary conclusions can be drawn:

i) $\tilde{v}_{eff}$ presents a nontrivial dependence on the field
amplitudes: in addition to appearing as explicit factors in the above
expression, $B_{1}$ and $B_{2}$ enter the arguments of the Bessel
functions. Actually, because of the form of those functions, there
is no monotonous increase of the effective interaction strength with
the amplitudes of the field components. Moreover, outside the limited
range of arguments where the applied perturbative treatment is valid,
a complex dependence of the effective coupling term on $B_{1}$ and
$B_{2}$ can be expected. 

It is also worth stressing the relevance of the Franck-Condon factor
$F(p)$, {[}see Eq. (3){]}, to the derived effective strength. $F(p)$,
which determines the magnitude of the original coupling, and, consequently,
the form of the dressed states, enters $v_{eff}$ via the product
of state projections $\left\langle M\right|\left.R\right\rangle \left\langle R\right|\left.A\right\rangle $.

ii) It is apparent that the conversion process can be controlled through
the choice of the phase difference $\chi$. Significant changes in
the effective interaction strength can be induced by varying $\chi$.
In particular, by taking $\chi=\pi/2$ ($-\pi/2$), a constructive
(destructive) interference is brought about. Furthermore, for $\chi=-\pi/2$,
a complete inhibition of the molecule generation process can be achieved
by properly choosing the values of the field amplitudes. Fig. 1 corresponds
to a set of parameters that exemplifies that situation. 

iii) From Eq. (17), one consistently recovers the results obtained
for a single-frequency driving field in \cite{key-BrouardAssoc}.
Specifically, by taking $B_{2}=0$, the result found for the main
resonance is reproduced. Alternatively, by fixing $B_{1}=0,$ we obtain
the renormalized coupling constant associated with the first subharmonic
resonance. In our framework, it is possible to obtain the field amplitude
$B_{1}$(or $B_{2}$) that optimizes the effective interaction strength
$\tilde{v}_{eff}$ for the main-resonance case (or for the first subharmonic.)
It is worth stressing the potential practical interest of the use
of subharmonic frequencies: they can allow the access to binding energies
that are outside the available bandwidth of the current techniques
for  generating magnetic fields.

iv) The additional elements of control introduced in the molecule-association
setup configure a more versatile scenario. In it, a variety of behaviors
can be implemented, for instance, via sudden or adiabatic changes
of the set of field parameters. Similarities with the typical arrangements
for Ramsey spectroscopy \cite{key-Ramsey-1,key-Donley,key-olsen}
can be traced: different frequencies can be apparent in the evolution
of the system by connecting (or disconnecting) the oscillating terms.
(Here, it is worth mentioning the interest of parallel setups, like
that studied in \cite{key-mark}.)

The above arguments are illustrated in Figs. 2 and 3. The case (a)
in Fig 2 exemplifies how significant changes in the magnitude of the
effective coupling constant can be achieved by varying the phase difference.
The previously mentioned suppression of the conversion process for
$\chi=-\pi/2$ and its intensification for $\chi=\pi/2$ are apparent.
Case (b), which corresponds to a different set of field amplitudes,
represents a more smooth variation of $\left|\tilde{v}_{eff}\right|$
with $\chi$. As can be shown from Eq. (17), the changes induced by
varying $\chi$ are more drastic as the magnitudes of the two terms
that contribute to $\tilde{v}_{eff}$ approach. Fig. 3 illustrates
the dependence of the interaction strength on $B_{2}$. The variety
of behaviors that can emerge depending on the system parameters is
again evident. Note that the eventual monotonous increase of $\left|\tilde{v}_{eff}\right|$
with $B_{2}$ corresponds to the regime where the role of the second
field is dominant. (Here, it is worth recalling the need of working
with relatively small arguments of the Bessel functions to guarantee
the applicability of the approach.) We emphasize that Eq. (17) allows
significant predictive power on the system dynamics; actually, the
possibility of using our proposal to control the conversion process
is apparent. \medskip{}

\includegraphics[scale=0.4]{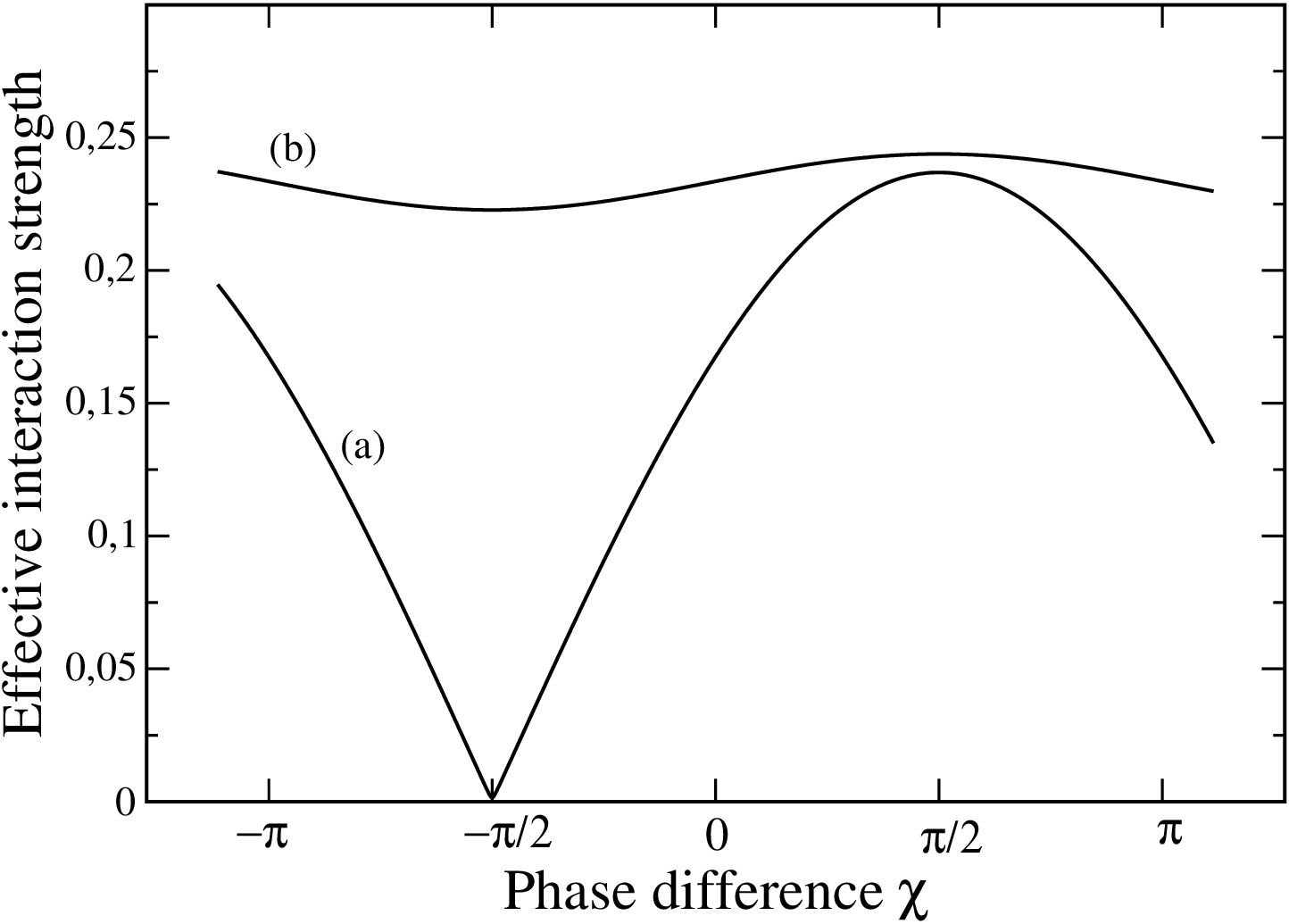}

\begin{figure}[H]
\caption{Magnitude of the effective interaction strength $\left|\tilde{v}_{eff}\right|$
as a function of the phase difference $\chi$ for two sets of magnetic-field
amplitudes: $B_{1}=0.255$, and $B_{2}=1.0$ (a); $B_{1}=0.5$, and
$B_{2}=0.3$ (b). In both cases, $\zeta_{1}=1.0$, and $\zeta_{2}=0.5$.
(The phase difference is expressed in radians; arbitrary units are
used for the rest of magnitudes.)}
\end{figure}

\medskip{}

\includegraphics[scale=0.4]{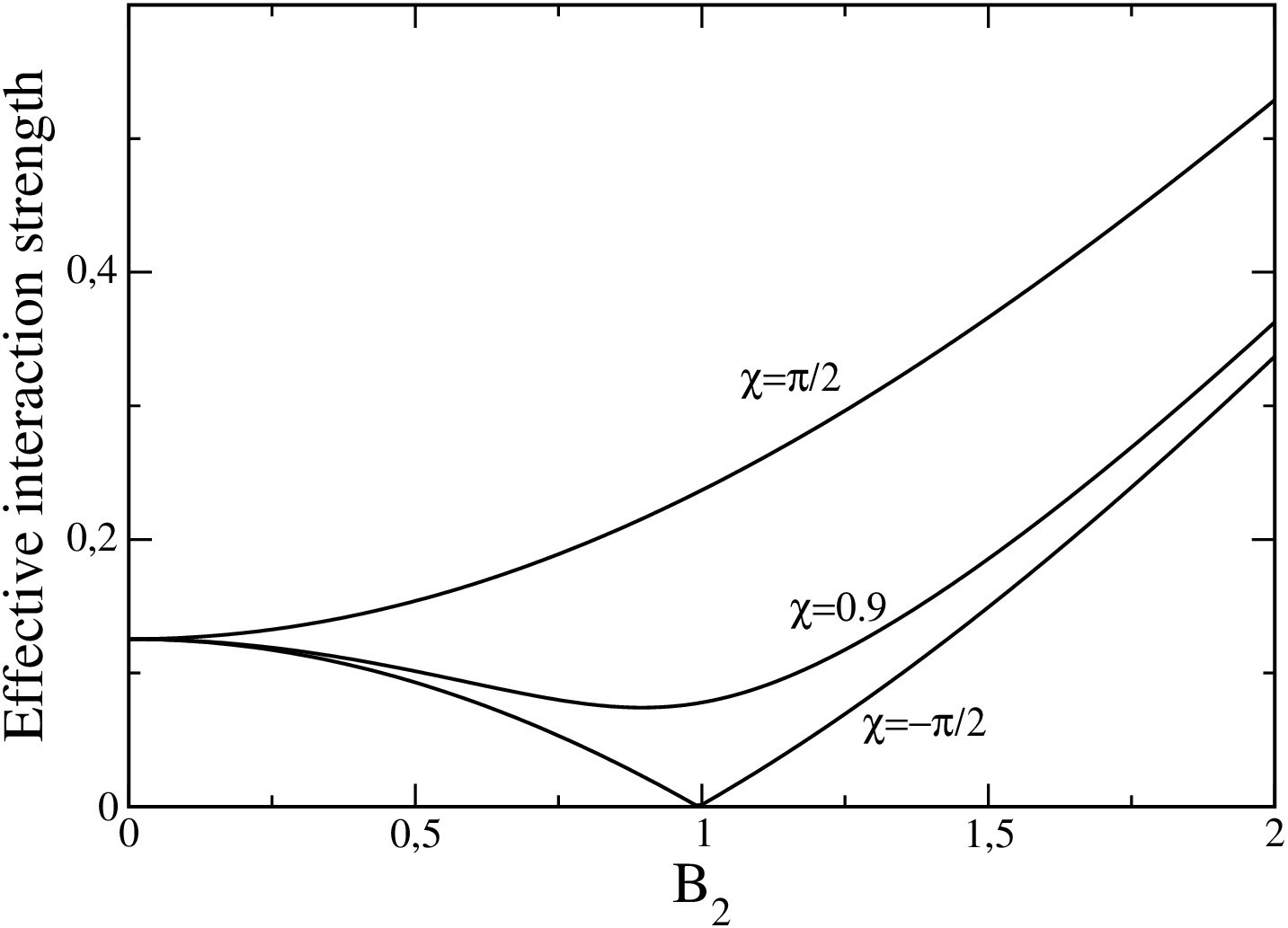}

\begin{figure}[H]
\caption{Magnitude of the effective interaction strength $\left|\tilde{v}_{eff}\right|$
as a function of the amplitude of the magnetic field $\vec{B_{2}}$
for three different values of the phase difference. ($B_{1}=0.255$,
and, as in Fig. 2, $\zeta_{1}=1.0$, and $\zeta_{2}=0.5$.) (The phase
difference is in radians; arbitrary units are used for the rest of
magnitudes.)}
\end{figure}

We stress that, in the present work, we focus on a primary aspect
of the dynamics that can be used to control the efficiency of the
atom-molecule conversion process. Namely, we concentrate on the variation
of the effective coupling constant that can be achieved with the proposed
arrangements of magnetic fields. Additional aspects of the dynamics,
which can limit  the conversion efficiency, were analyzed for the
case of single-frequency field in \cite{key-BrouardAssoc}. The conclusions
of that study are directly applicable to the present scenario.

\subsection{Applicability of the general approach to the analysis of field-quadratic
terms in the binding energy of the FR state}

By now we have considered a perturbative regime for the modulation
field. Correspondingly, we have neglected second-order effects in
the variation of the original detuning $\epsilon_{B_{0}}$ induced
by $B_{m}(t)$. However, since those effects have been observed in
some of the experimental realizations of the system, it is worth evaluating
here their implications to the dynamics. In this sense, we must recall
the results of \cite{key-Inguscio}, which uncovered the existence
of a shift in the molecular energy proportional to the square amplitude
of the modulation field. (In that case, a single-frequency driving
field was applied.) Actually, that energy displacement was found to
significantly modify the resonance condition. We will see that the
approach presented above can be adapted to account for effects rooted
in terms quadratic in the field $B_{m}(t)$. Moreover, we will show
that the combined effects of first and second-order terms in $B_{m}(t)$
can be regarded as a particular case of the interference scheme previously
developed. Convenient for the simplicity of the presentation is to
carry out the analysis with a single-frequency driving field, (as
corresponds to the experiments of \cite{key-Inguscio}.) The generalization
of the study to the case of a biharmonic driving will be discussed. 

Hence, we assume now that the magnetic-field modulation modifies the
undriven Hamiltonian in the form $H_{1}\rightarrow H_{2}=H_{1}+\left[C_{B_{0}}^{(1)}B_{m}(t)+C_{B_{0}}^{(2)}B_{m}^{2}(t)\right]\left|R\right\rangle \left\langle R\right|$,
where $C_{B_{0}}^{(1)}=\left[\frac{\partial\epsilon}{\partial B}\right]_{B_{0}}$corresponds
to the coefficient previously denoted as $C_{B_{0}}$, and we have
introduced $C_{B_{0}}^{(2)}=\frac{1}{2}\left[\frac{\partial^{2}\epsilon}{\partial B^{2}}\right]_{B_{0}}$
to characterize the (additional) quadratic dependence of the binding
energy on the magnetic field. Moreover, we take $B_{m}(t)=B_{1}\sin(\omega_{1}t)$,
and rewrite $H_{2}$ as 

\[
H_{2}=H_{1}+\left[\frac{1}{2}C_{B_{0}}^{(2)}B_{1}^{2}+C_{B_{0}}^{(1)}B_{1}\sin(\omega_{1}t)+\frac{1}{2}C_{B_{0}}^{(2)}B_{1}^{2}\sin(2\omega_{1}t-\pi/2)\right]\left|R\right\rangle \left\langle R\right|,
\]
which, in the basis of dressed states, (i.e., of eigenstates of $H_{1}$),
is cast into the form given by Eq. (5). Namely, it is expressed as 

\begin{eqnarray}
H_{2} & = & \left[-\hat{\epsilon}_{M}+\eta_{M}\hat{B}_{m}(t)\right]\left|M\right\rangle \left\langle M\right|+\left[\hat{E}_{A}+\eta_{A}(E_{A})\hat{B}_{m}(t)\right]\left|A\right\rangle \left\langle A\right|+\nonumber \\
 &  & \left[v{}_{eff}(E_{A})\hat{B}_{m}(t)\left|M\right\rangle \left\langle A\right|+\textrm{h.c.}\right],
\end{eqnarray}
where we have introduced the displaced energies $\hat{\epsilon}_{M}$
and $\hat{E}_{A}$ given by 

\begin{eqnarray}
\hat{\epsilon}_{M} & = & \epsilon_{M}-\eta_{M}\frac{1}{2}\frac{C_{B_{0}}^{(2)}}{C_{B_{0}}^{(1)}}B_{1}^{2},\\
\hat{E}_{A} & = & E_{A}+\eta_{A}(E_{A})\frac{1}{2}\frac{C_{B_{0}}^{(2)}}{C_{B_{0}}^{(1)}}B_{1}^{2},
\end{eqnarray}
and the effective biharmonic modulation field $\hat{B}_{m}(t)$ defined
as 

\begin{equation}
\hat{B}_{m}(t)=B_{1}\sin(\omega_{1}t)+\frac{1}{2}\frac{C_{B_{0}}^{(2)}}{C_{B_{0}}^{(1)}}B_{1}^{2}\sin(2\omega_{1}t-\pi/2).
\end{equation}
Moreover, $\eta_{M}$, $\eta_{A}(E_{A})$, and $v{}_{eff}(E_{A})$
are respectively given by Eqs (6), (7), and (8), where we only make
the change of notation $C_{B_{0}}\rightarrow C_{B_{0}}^{(1)}$. 

It is then apparent that the general approach introduced to deal with
the interference of association paths can be applied to analyze the
effect of incorporating field-quadratic terms  into the bare-state
detuning. Notice that the effective field $\hat{B}_{m}(t)$ corresponds
to a particular case of the generic biharmonic field given by Eq.
(9) with $B_{2}=\frac{1}{2}\frac{C_{B_{0}}^{(2)}}{C_{B_{0}}^{(1)}}B_{1}^{2}$,
$\omega_{2}=2\omega_{1}$, and $\chi=-\pi/2$. Now, following the
procedure introduced in the general approach, we apply the appropriate
unitary transformation (with $\omega=\omega_{1}$), {[}see Eq. (10){]},
and identify the dominant secular terms in the transformed Hamiltonian.
Accordingly, the reduced Hamiltonian is found to be given by 

\begin{eqnarray}
\hat{H}_{eff} & = & \hat{E}_{M}\left|M\right\rangle \left\langle M\right|+\hat{E}_{A}\left|A\right\rangle \left\langle A\right|+\nonumber \\
 &  & \left[\hat{v}_{eff}(E_{A})\left|M\right\rangle \left\langle A\right|+\textrm{h.c.}\right],
\end{eqnarray}
where $\hat{E}_{M}=\omega_{1}-\hat{\epsilon}_{M}$, and 

\[
\hat{v}_{eff}=J_{0}(\zeta_{2}B_{2})\left[-iJ_{0}(\zeta_{1}B_{1})B_{1}+J_{1}(\zeta_{1}B_{1})B_{2}\right]\frac{v_{eff}}{2}.
\]
Taking into account the expressions of the Bessel functions in the
limit of small arguments and keeping only terms till second order
in $B_{1}$, we can make the additional approximation for the effective
interaction strength

\[
\hat{v}_{eff}=-iJ_{0}(\zeta_{1}B_{1})B_{1}\frac{v_{eff}}{2}.
\]
This expression corresponds to that obtained when only linear terms
are considered. Hence, at the considered level of approximation, the
quadratic corrections affect only the splitting between the effective
atomic and molecular levels, which now reads $\hat{\Delta}\equiv\hat{E}_{M}-\hat{E}_{A}=\Delta-(\eta_{M}-\eta_{A})\frac{1}{2}\frac{C_{B_{0}}^{(2)}}{C_{B_{0}}^{(1)}}B_{1}^{2}$.
Consequently, the resonance frequency is given by $\omega_{1}^{(R)}=\epsilon_{M}+E_{A}+(\eta_{M}-\eta_{A})\frac{1}{2}\frac{C_{B_{0}}^{(2)}}{C_{B_{0}}^{(1)}}B_{1}^{2}$.
In agreement with the findings of \cite{key-Inguscio}, we have obtained
a displacement of the resonance which is quadratic with the field
amplitude. The magnitude of this shift depends on the (linear and
quadratic) coefficients that characterize the field dependence of
the detuning, and, also, on the projections between the FR state $\left|R\right\rangle $
and the dressed states $\left|M\right\rangle $ and $\left|A\right\rangle $
involved in the transition. 

It is apparent that, in the case of considering a biharmonic function
for $B_{m}(t)$, the quadratic shift of the resonance is still present.
Since additional oscillating terms appear in the effective field $\hat{B}_{m}(t)$,
a more complex expression is obtained for $\hat{v}_{eff}$.

\section{Discussion of the relevance of the different sources of decoherence }

The feasibility of controlling the interference between association
paths depends crucially on the coherent response of the system to
the modulation field. In the above description of the scheme, we deal
with a simplified scenario where coherence was assumed. Now, we turn
to analyze the robustness of the proposal against different components
of the system which can affect the coherent character of the dynamics. 

First, we recall that the atomic state involved in the transition
has been assumed to be effectively isolated from the rest of the atomic
set. Accordingly,  Eq. (11) defines an effective (linear) Rabi model
involving $\left|M\right\rangle $ and an (isolated) generic state
$\left|A\right\rangle $. The quasi-continuum atomic structure, rooted
in the confinement, can be incorporated into the model by describing
the atomic set in terms of the density of states and of the coupling
function with the bound state. Different positions of $\left|M\right\rangle $
with respect to the atomic threshold can be realized by changing the
modulation characteristics. For the different cases, the damping and
energy shift of the discrete state are evaluated \cite{key-CohenAvan}.
The results allow the characterization of the time scale where, despite
the existence of damping, a coherent response can be assumed. In that
scale, the effective discretization, which requires the modification
of the coupling constant to incorporate the properties of the density
of states and the confining volume \cite{key-Mies}, is sound. One
should take into account that, in this approach, the variation of
the energy $E_{A}$ along the atomic set is still contemplated. This
is particularly important to describe experiments done on uncondensed
samples \cite{key-ThomsonAssoc,key-Hanna Assoc}, where the starting
point is a thermal distribution of atoms. The variation of $E_{A}$
in the dominion of atomic energies implies dealing with a continuous
range of frequencies in the system response, and therefore, with the
emergence of dephasing effects when the averaging over the distribution
is carried out. The characteristic time for the associated decoherence
is determined by the width of the state distribution, and, therefore,
by the temperature.

A second line of generalization of our model refers to the inclusion
of many-body effects. An operative reduction of the complete microscopic
dynamics must incorporate two fundamental elements. First, for a condensate
preparation, since, each of the $N$ identical atoms of the system
can interact with $N-\textrm{1}\simeq N$ others to form the molecular
species, a scaling of the interaction strength in the equations for
the evolution of the populations is needed. This \emph{bosonic stimulation}
can significantly affect the efficiency of the population transfer.
Second, the correct procedure to \emph{renormalize} the dynamical
equations must take into account that the number of single atoms in
the system changes as the conversion progresses. This implies the
inclusion of an effective interaction strength which depends on the
(evolving) atomic population, and, therefore, the use of nonlinear
equations. (The generalization along this line and the inclusion of
the dependence of the interaction strength on the confining volume
are crucial for understanding the relevance of the atom density to
the conversion efficiency.) Following these lines, our former (simplified)
description of the system is replaced by a\emph{ nonlinear }Rabi model
\cite{key-Timmermans,key-ReviewKohler,key-Mies}. 

Apart from many-body physics, the model must incorporate molecular
decay if the time scale for this process is smaller than that of the
complete conversion transfer. Losses can be incorporated in a phenomenological
way: they can be simply described as a depletion of the molecular
population characterized by a constant rate \cite{key-Inguscio,key-BrouardAssoc}.
This implies dealing with a damped \emph{nonlinear }Rabi model. 

Previous reports on atom-molecule conversion methods have stressed
the technical difficulties of avoiding magnetic non-uniformity in
the practical setups \cite{key-olsen}. Hence, in addition to damping
due to dissociation and thermal decoherence, it is worth considering
the spatial variations in the applied magnetic field as a possible
source of dephasing. Our approach provides a framework where the inhomogeneities
can be straightforwardly incorporated: in a local-density approximation,
a nonuniform magnetic field can be tackled by simply averaging the
results previously obtained for the homogeneous case over the distribution
of field values.  

Obviously, the above elements are not specific to the proposed setup.
They are present in the standard scenario defined by a single-frequency
modulation field, where, despite their effects, there is a significant
time scale where a coherent response is observed. Then, we can infer
that, although those components can affect the system dynamics, they
do not preclude the applicability of the interference scheme. 

A source of decoherence specific to the interference scheme is that
rooted in field-phase fluctuations. The precise fixing of the phase
difference between the modulation-field components is essential for
the method to work. We can incorporate phase noise into our framework
through a distribution centered on the optimum phase value and with
a width depending on the particular setup conditions. In our reduced
picture, the evolution of the atomic and molecular populations corresponds
to a Rabi oscillation, (linear or nonlinear, depending on the magnitude
of many-body effects), with the frequency being determined by the
effective interaction strength and by the detuning. Then, fluctuations
in the phase lead to the random variation of the frequency, and, consequently,
to dephasing in the output oscillations. The time scale where coherence
persists depends then on the width of the noise distribution. Again,
it seems that, in standard arrangements, reduced levels of phase noise
can be realized.

\section{Concluding remarks}

We have shown that the interference between association paths can
provide a tool for controlling the conversion of ultracold atoms into
molecules in the magnetic-field modulation scenario. The coherence
required for the interference scheme to work seems to be feasible
in standard arrangements. Actually, although the simplified picture
used to present the method of control can be considerably modified
when different components of the system, (e.g., many-body effects,
molecular decay, decoherence, or nonuniformities in the field) are
incorporated, the central effects generated by the interference persist.
Namely, the induced modification of the interaction strength is present
in any required generalization of the applied approach. The applicability
of the study to a system of fermions, in particular, the differential
treatment of many-body effects, will be the subject of future work.

The second main goal of our study has been the characterization of
second-order effects in the field modulation. We have incorporated
into the basic model the quadratic dependence of the binding energy
of the FR state on the driving field. We have shown that the combined
effects of linear and quadratic terms can be regarded as a particular
case of the proposed interference scheme. Actually, those effects
have been described using the general approach previously developed
to deal with control issues. As a result, we have explained the shift
in the resonance frequency induced by the field quadratic terms, observed
in previous experimental work.

\end{document}